\documentclass[12pt]{article}
\newcommand{\be}{\begin{equation}}
\newcommand{\ee}{\end{equation}}
\newcommand{\bea}{\begin{eqnarray}}
\newcommand{\eea}{\end{eqnarray}}

\newcommand{\ads}{AdS/CFT~}
\def\({\left(} \def\){\right)}
\usepackage{amssymb}
\usepackage{epsfig}
\usepackage{graphicx}
\usepackage{amsmath,amssymb,graphicx,mathrsfs}
\usepackage{hyperref}
\begin{document}

\title{\vspace{-1.7in} The Noether charge entropy in anti-deSitter space and its field theory dual}
\author{\large  Ram Brustein \\   Department of Physics, Ben-Gurion
University,\\
    Beer-Sheva, 84105 Israel, E-mail: ramyb@bgu.ac.il\\ \\
Dan Gorbonos \\ Department of Physics, University of Alberta,
\\ Edmonton, Alberta, Canada T6G 2G7 \\
    E-mail: gorbonos@phys.ualberta.ca}
\date{}
\maketitle

\abstract{

We express the Noether charge entropy density of a black brane in
anti-deSitter space in terms of local operators in the anti-deSitter
space bulk. We find that Wald's expression for the Noether charge
entropy needs to be modified away from the horizon by an additional
term that vanishes on the horizon.  We then determine the field
theory dual of the Noether charge entropy for theories that
asymptote to Einstein theory. We do so by calculating the value of
the entropy density at the anti-deSitter space boundary and applying
the standard rules of the \ads correspondence. We interpret the
variation of the entropy density operator from the horizon to the boundary as
due to the renormalization of the effective gravitational couplings
as they flow from the ultra-violet to the infra-red. We discuss the
cases of Einstein-Hilbert theory and $f(R)$ theories in detail and
make general comments about more complicated cases. }

\maketitle

\pagebreak
\tableofcontents

\section{Introduction}
\label{intro}

The Noether charge entropy (NCE) of black holes has been proposed by
Wald~(\cite{wald1,wald2}) and was expressed  as
\begin{equation}
\label{waldentropygen}
S_{W}=-2 \pi \oint\limits_{\Sigma}
\left(\frac{\delta\mathscr{L}}{\delta R_{abcd}}\right)^{\!\!(0)}
\hat\epsilon_{ab}\epsilon_{cd}
\end{equation}
in \cite{jacobson}. The NCE  will be discussed in detail in Section~\ref{localentropy}.

It was shown in~\cite{first} that the
kinetic terms for metric perturbations
$
g_{\mu\nu}=\bar{g}_{\mu\nu}+h_{\mu\nu}
$
for the general action (\ref{genlag}) are given by
\begin{equation}
\label{final}
\delta I^{(2)}= \int\!\! d^{d+1} x\sqrt{-\overline{g}}\ \frac{1}{2}
\left(\frac{{\delta \mathscr{L}}}{ \delta
R_{\rho\mu\lambda\nu}}\right)^{\!\!(0)}
\left(\overline{\nabla}_{\delta}h_{\lambda\mu}\overline{\nabla}^{\delta}h_{\nu\rho}
+2\overline{\nabla}^{\delta}h_{\lambda\rho}\overline{\nabla}_{\mu}h_{\nu\delta}
\right).
\end{equation}
The background covariant derivative is denoted by $\overline{\nabla}$ and the superscript $(0)$ indicates that the partial derivative
$\left(\frac{\delta\mathscr{L}}{\delta R_{abcd}}\right)^{\!\!(0)}$
is evaluated on the solution of the equations of motion. In this expansion we keep only terms
that contain two factors of the metric perturbation
and two background covariant derivatives. The coefficient tensor
$\left(\frac{\delta\mathscr{L}}{\delta R_{abcd}}\right)^{\!\!(0)}$ determines the various
effective gravitational coupling constants for the different polarizations.
For example, the effective coupling
relevant to the entropy is defined as
\begin{equation}
\label{kappadefgeneral}
\frac{1}{\left(\kappa_{eff}\right)^2} = -\frac{1}{4}\left( \frac{\delta\mathscr{L}}{\delta R_{abcd}}\right)^{\!\!(0)} \hat\epsilon_{ab}\hat\epsilon_{cd},
\end{equation}
so that Wald's entropy (\ref{waldentropygen}) can be written as
\begin{equation}
\label{waldareaI}
S_{W}= \frac{1}{4}\oint\limits_{\Sigma}\frac{8 \pi}{\left(\kappa_{eff}\right)^2}\,dA,
\end{equation}
$dA$ being the surface element. The binormal vectors $\hat\epsilon_{ab}$, in this case, pick a specific polarization
of the metric fluctuations that corresponds to  fluctuations of  the area of the bifurcation surface $\Sigma$.

The effective gravitational coupling constant  $\kappa_{eff}$ appears to have two roles -- it determines the coupling constant for a specific polarization and it also determines the entropy density per unit area on the bifurcate surface $\Sigma$. Considering $\kappa_{eff}$ as a coupling constant leads us  to look at the coefficients tensor $\left( \frac{\delta\mathscr{L}}{\delta R_{abcd}}\right)^{\!\!(0)}$ away from the horizon.

The gravitational coupling constant gives the bulk entropy density
evaluated on the horizon.  Our interpretation of the entropy density
as a coupling constant highlights the fact that, from this point of
view, the entropy is given in terms of a local quantity in the bulk.
Yet, there is another description of the entropy density in the
context of \ads as the entropy density of the field theory on the
boundary. According to the \ads correspondence the two descriptions
should agree. We propose that the coupling constant should be promoted to a
field in the bulk by allowing it to vary in the radial direction.
Doing so will allow us to use the standard tools of the \ads
correspondence in order to relate it to a dual operator in the field theory.
We will then be able to relate the entropy of the black brane to the
entropy of the boundary field theory. In addition we would
like to look at the renormalization group (RG) flow of the effective
coupling constant from the horizon (UV) to the boundary (IR) in
asymptotically AdS spacetimes.

In this paper we consider the following ansatz for the metric of a
d+1-dimensional black brane in AdS
\begin{equation}
ds^{2}=-g_{tt}\,dt^{2}+g_{rr}\,dr^{2}+g_{xx}\,dx^{i}\,dx_{i}.
\label{brane}
\end{equation}
The black brane event horizon is at $r=r_h$, where $g_{tt}$ has a
first order zero and $g_{rr}$ has a first order pole. We assume that
all other metric components are finite at the horizon. All the
metric components are taken to depend only on $r$ and therefore the
metric is Poincare invariant in the $(t,x_i)$ subspace. We assume
that the $AdS$ boundary is at $r=\infty$. We also assume that
asymptotic form of the metric approaches the AdS metric. Any
asymptotically AdS metric
can be brought to the Fefferman-Graham form near the
boundary~\cite{Graham}
\begin{equation}
\label{feffgraham}
ds^{2}=l^{2}\,\(\frac{d\,\rho^{2}}{4\,\rho^{2}}+ \frac{1}{\rho}\,g_{ij}(x,\rho)\,dx^{i}dx^{j}\),
\end{equation}
where
\begin{equation}
g(x,\rho)=g_{(0)}+...+\rho^{\frac{d}{2}}g_{(d)}+ h_{(d)}\,\rho^{\frac{d}{2}}\,\log{\rho}+...,
\end{equation}
$l$ is related to the cosmological constant as
$\Lambda=-\frac{d\,(d-1)}{2\,l^{2}}$ and the boundary is at
$\rho=0$ (The coefficient $h_{(d)}$ is present only when $d$
is even).

We consider a general theory of gravity whose action depends on
the metric $g_{\mu\nu}$, the curvature (through the Riemann tensor)
and on matter fields $\phi$ and their covariant derivatives
\begin{equation}
\label{genlag} I=\int\!\! d^{d+1} x \sqrt{-g}\
\mathscr{L}\left(R_{\rho\mu\lambda\nu},g_{\mu\nu},\nabla_\sigma
R_{\rho\mu\lambda\nu},\phi,\nabla\phi,\ldots\right).
\end{equation}

We assume that the Lagrangian~(\ref{genlag}) has stationary black
brane solutions of the form (\ref{brane}) with a bifurcate Killing
horizon. In this paper we will consider only higher-derivative
actions such that the effective gravitational coupling at the
boundary approaches asymptotically the Newton's constant of the
Einstein-Hilbert Lagrangian. We will discuss this requirement in
section~\ref{the dual}.

\section{The local form of the Noether charge entropy density}
\label{localentropy}

The NCE  is
given by
\begin{equation}
S_{W}=-2 \pi \oint\limits_{\Sigma}
\left(\frac{\delta\mathscr{L}}{\delta R_{abcd}}\right)^{\!\!(0)}
\hat\epsilon_{ab}\epsilon_{cd}.
\end{equation}
The variation of the Lagrangian with respect to $R_{abcd}$ is performed
as if $R_{abcd}$ and the metric $g_{mn}$ are independent.

The surface element $(d-1)$-form $\epsilon_{cd}$ is defined on the
space-like bifurcation surface $\Sigma$. The hatted variable
$\hat\epsilon_{ab}$ is the binormal vector to the bifurcation
surface defined as
$\hat\epsilon_{ab}=\nabla_{a}\widetilde{\chi}_{b}$, the binormal is antisymmetric under the exchange $a\leftrightarrow
b$. The vector
$\widetilde{\chi}^{b}$ is the normalized Killing vector which
generates the Killing horizon. The Killing vector $\chi^{b}$ satisfies
the Killing equation
\begin{equation}
\nabla_{a}\chi_{b}\,\nabla^{a}\chi^{b}=-2\,\kappa^{2},
\end{equation}
$\kappa$ being is the surface gravity. The normalized Killing vector is
defined so that $\chi^{b}=\kappa\widetilde{\chi}^{b}$. Due to this
normalization the entropy is computed in units such that the brane
temperature is equal to $\frac{1}{2\,\pi}$. If this normalization is not enforced then the
entropy formula reads
\begin{equation}
\label{waldentropygen1} S_{W}=-\frac{1}{T} \oint\limits_{\Sigma}
\left(\frac{\delta\mathscr{L}}{\delta R_{abcd}}\right)^{\!\!(0)}
\nabla_{a}\chi_{b}\epsilon_{cd}.
\end{equation}

Wald's entropy for a black brane solution of the form (\ref{brane}) is proportional to $A_h$,
the area of the horizon at $r=r_h$,
\begin{equation}
S_{W}=- 8\,\pi A_{h} \,\left. \frac{\partial \mathscr{L}}{\partial R^{rt}_{\phantom{rt}rt}}\right|_{r=r_h}.
\label{wald for static}
\end{equation}
We can regard
$
\Theta^t_{\phantom{t}t}(r)\equiv -2 \frac{\partial \mathscr{L}}{\partial R^{rt}_{\phantom{rt}rt}}
$
as a function of the radial coordinate
$r$ whose value at the horizon determines the entropy density.
Let us define another quantity $\Theta^x_{\phantom{x}x}(r)$,
\begin{eqnarray}
\Theta^t_{\phantom{t}t}(r)&\equiv&-2\frac{\partial \mathscr{L}}{\partial R^{rt}_{\phantom{rt}rt}},\\
\Theta^x_{\phantom{x}x}(r)&\equiv&2\frac{\partial
\mathscr{L}}{\partial R^{rx}_{\phantom{rx}rx}}.
\end{eqnarray}
The motivations for these definitions will be revealed and explained later.

With the newly defined operators we may define quantities that will be closely related to thermodynamic quantities in the field theory. First we define
\begin{equation}
\sigma_t \equiv 2\,
\Theta^t_{\phantom{t}t}(r)(g_{xx})^{\frac{d-1}{2}}. \label{defeps}
\end{equation}
The entropy per unit area on the horizon is then related to $\sigma_t$
\begin{equation}
s T=\sigma_t|_{r=r_h}.
\end{equation}
Additionally, the metric is translation invariant in the non-radial spatial directions. Hence, using the Killing vectors in
these directions one can define an analogous quantity associated
with $\Theta^x_{\phantom{x}x}(r)$,
\begin{equation}
\sigma_x \equiv 2\,\Theta^x_{\phantom{x}x}(r)(g_{xx})^{\frac{d-2
}{2}}\sqrt{-g_{tt}}. \label{defp}
\end{equation}
In Eq.~(\ref{defp}) we have used the binormal $\hat\epsilon_{rx}$
with respect to the radial direction and one of the spatial
orthogonal directions. Clearly, $\sigma_x$ vanishes at the horizon
due to the vanishing of $g_{tt}$ there. Nevertheless, we will show
that $\sigma_x$ has a definite meaning from the field theory  point
of view. The vanishing of $\sigma_x$  at the horizon can be viewed
as due to being infinitely red-shifted there. As a precursor to the
later discussion let us notice that on the horizon
$\sigma_t+\sigma_x=s T$. For later reference we will also define a
linear combination of $\Theta^t_{\phantom{t}t}$ and
$\Theta^x_{\phantom{x}x}$,
\begin{equation}
{\cal S} \equiv-\frac{1}{2}\(\Theta^{t}_{\phantom{t}t}+\Theta^{x}_{\phantom{x}x}\). \label{eps} \\
\end{equation}

As stated previously,  we will consider only higher derivative
actions with asymptotically AdS solutions, for which the effective
gravitational couplings approach their values in the
Einstein-Hilbert Lagrangian, and its definition will be given in
section~\ref{the dual}. In particular,
\begin{equation}
\begin{aligned}
&\lim_{r \rightarrow \infty}\Theta^t_{\phantom{t}t}(r)=\frac{1}{16\,\pi\,G_{d+1}}   \quad,\\
 &\lim_{r \rightarrow
\infty}\Theta^x_{\phantom{x}x}(r)=\frac{1}{16\,\pi\,G_{d+1}}
\quad .
\end{aligned}
\end{equation}
Here $G_{d+1}$ is the $d+1$-dimensional Newton's constant.

\section{The Noether charge entropy density in the bulk and on the boundary}

\subsection{Conjugate variables and Hamiltonian holography}

We would like to find the expectation value of the operator dual to
${\cal S}$ defined in Eq.~(\ref{eps}).
For this purpose let us recall  some general principles of the \ads
correspondence.

Let us look at a general action in the bulk of the form
\begin{equation}
S=\int^{\infty}_{r_h}\!\! dr\int\!\! dt\,\int\!\! d^{d-1}x {\cal
L}\(\Phi_{A}(r,x)\)
\end{equation}
where $\Phi_{A}(r,x)$ represents collectively all the fields of the
theory.

According to the \ads
correspondence and the prescription for holographic renormalization,
the expectation value of the  operator $O_{A}(x^{i})$ on the
boundary that is dual to the field $\Phi_{A}(r,x^{i})$ in the bulk is given by
\begin{equation}
\langle    O_{A}(x^{i})\rangle    \equiv \frac{1}{\sqrt{-h}}\,\frac{\delta
S_{on-shell,ren}}{\delta \Phi^{(0)}_{A}}.
\label{dualvev}
\end{equation}
$S_{on-shell,ren}$ is the renormalized on-shell action, $\Phi^{(0)}_{A}(x^{i})$ is the
value of the bulk field on the boundary (which is also the source
for the generating function of the field theory) and $\sqrt{-h}$ is the determinant of
boundary metric.

Let us consider a variation of the bulk action
\begin{eqnarray}
\delta S &=&\left.\frac{\partial\,{\cal
L}}{\partial\,\dot{\Phi}_{A}}\,\delta\,\Phi_{A}\right|^{\infty}_{r_h}
+\int^{\infty}_{r_h}\!\!\,dr\,\delta\,\Phi_{A}\,\left[\frac{\partial\,{\cal
L}}{\partial\,\Phi_{A}}-
\partial_{r}\,\(\frac{\partial\,{\cal L}}{\partial\,\dot{\Phi}_{A}}\)\right]
\cr &=& \left. \frac{\partial\,{\cal
L}}{\partial\,\dot{\Phi}_{A}}\,\delta\,\Phi_{A}\right|^{\infty}_{r_h}
\end{eqnarray}
(a dot denotes the differentiation with respect to the radial coordinate).
Recall that the AdS boundary is at $r\to \infty$  and the horizon is at $r=r_h$.

Introducing the conjugate momentum in the bulk
\begin{equation}
\Pi_{{\Phi}_{A}}(r,x^i)\equiv\frac{\partial\,{\cal
L}}{\partial\,\dot{\Phi}_{A}(r,x^i)}
\end{equation}
we can express the variation of the on-shell action  as
\begin{equation}
\delta S_{on-shell}=\lim_{r \rightarrow
\infty}\Pi_{{\Phi}_{A}}(r,x^i)\,\delta\,\Phi^{(0)}_{A}-\lim_{r
\rightarrow r_h}\Pi_{{\Phi}_{A}}(r,x^i)\,\delta\,\Phi_{A}(r_h).
\end{equation}

Then we can use Eq.~(\ref{dualvev}) to obtain the expectation value
of the dual operator from the value of the conjugate momentum on the
boundary after an appropriate renormalization (for details
see~\cite{papa}),
\begin{equation}
\langle    O_{A}(x^i)\rangle    =\lim_{r \rightarrow
\infty}\frac{1}{\sqrt{-h}}\Pi_{{\Phi}_{A}}(r,x^i)_{ren}.
\label{the momentum}
\end{equation}

\subsection{The decomposition of the action with respect to a radial hypersurface}

In this section  we decompose a general action with respect to the
hypersurface $r=constant$.
The action is  of the form
\begin{equation}
\label{theaction}
S = \frac{1}{16\,\pi\,G_{d+1}} \int\!\! dr\int\!\! dt\int\!\!\,d^{d-1}x
\,\sqrt{-g}\,{\cal L} (g_{a\,b},R_{a\, b \, c \, d}, \Phi_A, \nabla\Phi_A, \cdots).
\end{equation}
We wish to find the functional derivative  of the action with respect to the Riemann tensor  $\delta S / \delta R_{rbcr}$.
If derivatives of the Riemann tensor appear in ${\cal L}$ then one
has to perform integrations by parts first and then take the
derivative. The procedure is similar to finding the Euler-Lagrange
equations in a theory with higher derivatives of the canonical
variables.
If matter fields (collectively denoted by $\Phi_A$) and their covariant derivatives appear in ${\cal
L}$, the covariant derivatives  have to be expressed in
terms of the Riemann tensor prior to evaluation of the functional derivative \cite{wald2}. The remaining terms that contain matter fields that do not depend on the Riemann tensor will be decomposed in a  procedure similar to the one given below for the Riemann tensor. Such terms do not affect any of the results below. Therefore we can safely ignore them
for the sake of simplicity.

The decomposition that we will describe follows a procedure similar  to that introduced in~\cite{brown} for the decomposition with respect to
a constant $t$  hypersurface. The decomposition with respect to
$r=const.$ has been extensively used in the  \ads context since
the radial direction in many ways plays the role of time (see, for
example, \cite{papa}).

We define a hypersurface $r=const$. for the geometry given by
(\ref{brane}) by its normal $n_{r}=-\frac{1}{\sqrt{g_{rr}}}$. Let
$e^{a}_{\ \alpha}$ be a basis of tangent vectors to the
hypersurface, and $h_{\alpha\beta}=g_{ab}\,e^{a}_{\
\alpha}\,e^{b}_{\ \beta}$ the induced metric of the hypersurface. We
use Greek indices to denote the induced coordinates on the
hypersurface. The decomposition of the metric $g_{ab}$ is then
\[g_{ab}=e^{\alpha}_{\ a}\,e^{\beta}_{\ b}\,h_{\alpha\beta}-n_{a}\,n_{b}.\]
The covariant derivative on the hypersurface is defined as a
projection of the general covariant derivative $D_{\alpha}\equiv
e^{a}_{\ \alpha}\,\nabla_{a}$.

Next we wish to decompose the Riemann tensor with respect to the
hypersurface as the first step towards a decomposition of a general
action. For this purpose we introduce the Lie derivative in the
direction of $n^{a}$, $L_{n}=\frac{1}{N} \partial_{r}$
where $N=\sqrt{g_{rr}}$ is the lapse function. We fix the shift
to be zero, so that the bulk metric has the following form
\begin{equation}
ds^{2}=N^{2}\,dr^{2}+h_{\alpha\beta}(r)\,dx^{\alpha}\,dx^{\beta}.
\label{metric:form}
\end{equation}

The Gauss-Codazzi-Ricci decomposition of $R_{abcd}$ is given then by
\begin{eqnarray}
e^{d}_{\ \delta}\,e^{c}_{\ \gamma}\,e^{b}_{\ \beta}\,e^{a}_{\ \alpha}\,R_{abcd}&=& R_{\alpha\beta\gamma\delta}+K_{\alpha\gamma}K_{\delta\beta}-K_{\alpha\delta}K_{\gamma\beta} \label{Riem1}\\
n^{a}e^{d}_{\ \delta}\,e^{c}_{\ \gamma}\,e^{b}_{\ \beta}\,R_{abcd} &=&D_{\gamma}K_{\beta\delta}-D_{\delta}K_{\beta\gamma} \label{Riem2}   \\
n^{a}n^{d}e_{\ \beta}^{b}\,e_{\ \gamma}^{c}\,R_{abcd}&=&L_n\,K_{\beta\gamma}+ K_{\beta\eta}K^{\eta}_{\phantom{\eta}\gamma}+\frac{D_{\beta}D_{\gamma}N}{N}.
\label{Riem3}
\end{eqnarray}

Now for the decomposition of the action let us introduce two
auxiliary non-dynamical tensors $V_{abcd}$ and $U_{abcd}$ and look
at the following action
\begin{eqnarray}
\widetilde{S}&=&\frac{1}{16\,\pi\,G_{d+1}}\int\!\! dr
\int\!\! dt \int\!\!\,d^{d-1}x\,\sqrt{-g}\ \times \cr && \,\left[{\cal L}
(g_{ab},V_{abcd})+U_ {abcd}\,R^{abcd}-U_ {abcd}\,V^{abcd}\right],
\end{eqnarray}
which is equivalent to the original action (\ref{theaction}) when we
substitute the equations of motion
\begin{eqnarray}
\frac{\partial  L}{\partial V_{abcd}}&=&U^{abcd},\\V^{abcd}&=&R^{abcd}.
\end{eqnarray}

\subsection{The dual of $\frac{\partial  L}{\partial R_{rbcr}}$}

With the auxiliary fields only one term in the action
$\widetilde{S}$ depends explicitly on $R_{abcd}$. We have to
decompose this term according to Eqs.~(\ref{Riem1})-(\ref{Riem3}),
\begin{eqnarray}
&&\widetilde{S}=\frac{1}{16\,\pi\,G_{d+1}}\int\!\! dr \int\!\! dt \int\!\! d^{d-1}x
\sqrt{-h}\,N\ \times \cr && \Biggl[\,U^{\alpha\beta\gamma\delta}\(R_{\alpha\beta\gamma\delta}+ 2\,K_{\alpha\gamma}K_{\beta\delta}\) +  8 U^{r\beta\gamma\delta }D_{\gamma}K_{\beta\delta}
\cr && + 4U^{r\beta\gamma r}\,N^{-1}\(\dot{K}_{\beta\gamma} + N\,K_{\beta\eta}K^{\eta}_{\phantom{\eta}\gamma}+D_{\beta}D_{\gamma}N\)+\cdots
\Biggr].
\label{stilde}
\end{eqnarray}
The dots denote the decomposition of the rest of the fields.

Since $K_{\beta\gamma}=\dot{h}_{\beta\gamma}/(2\,N)$,  the term
with $\dot{K}_{\beta\gamma}$ in Eq.~(\ref{stilde}) contains second
derivatives with respect to $r$. Following~\cite{brown} we introduce
an additional auxiliary field $P^{\alpha\beta}$ and treat
$K_{\alpha\beta}$ as an independent variable,
\begin{equation}
\begin{aligned}
&& \widetilde{S}=\frac{1}{16\,\pi\,G_{d+1}}\int\!\! dr \int\!\! dt
\int\!\! d^{d-1}x\,\,\Biggl\{
P^{\alpha\beta}\(2\,N\,K_{\alpha\beta}-\dot{h}_{\alpha\beta}\)  \cr
&&
+\sqrt{-h}\,N\Biggl[U^{\alpha\beta\gamma\delta}\(R_{\alpha\beta\gamma\delta}
+2\,K_{\alpha\gamma}K_{\beta\delta}\)
+8U^{r\beta\gamma\delta}D_{\gamma}K_{\beta\delta} \cr && +
4U^{r\beta\gamma r}\,N^{-1}\(\dot{K}_{\beta\gamma}
+N\,K_{\beta\eta}K^{\eta}_{\phantom{\eta}\gamma}+D_{\beta}D_{\gamma}N\)+\cdots
\Biggr]\Biggr\}. \label{action:Conjugate}
\end{aligned}
\end{equation}
Integrating by parts the term $ 1/{4\,\pi\,G_{d+1}}\ \sqrt{-h}\, U^{r\beta\gamma
r}\dot{K}_{\beta\gamma}$ in
Eq.~(\ref{action:Conjugate}) we find that the conjugate variables to
\begin{eqnarray}
\Theta^{tt} &\equiv& 2\,U^{r t t r}(r,t,x^{i}) \cr
\Theta^{xx} &\equiv& -2\,U^{r x x r}(r,t,x^{i}),
\end{eqnarray}
are
\begin{eqnarray}
\Pi_{\Theta^{tt}}(r,t,x^{i})&=&-\frac{\sqrt{-h}}{8\,\pi\,G_{d+1}}\,K_{tt}(r,t,x^{i})
\cr
\Pi_{\Theta^{xx}}(r,t,x^{i})&=&\frac{\sqrt{-h}}{8\,\pi\,G_{d+1}}\,K_{xx}(r,t,x^{i})\label{theconjugate}.
\end{eqnarray}

Then according to Eq.~(\ref{the momentum}) the one-point functions
of the dual operators to $\Theta^{t}_{\phantom{t}t}$ and
$\Theta^{x}_{\phantom{x}x}$ are given by
\begin{eqnarray}
\langle    O_{\Theta^{t}_{\phantom{t}t}}\rangle   & =& -\lim_{r \rightarrow \infty}\frac{\(K_{\phantom{t}t}^{t}\)_{ren}}{8\,\pi\,G_{d+1}}
\cr
\langle    O_{\Theta^{x}_{\phantom{x}x}}\rangle   & = &\lim_{r \rightarrow
\infty}\frac{\(K_{\phantom{x}x}^{x}\)_{ren}}{8\,\pi\,G_{d+1}}.
\end{eqnarray}

\section{The field theory dual of the Noether charge entropy density}
\label{the dual}

To complete the process of identifying the dual of the NCE density
we need to express the extrinsic curvature in terms of field theory
operators. In this paper we consider only higher-derivative actions
with solutions whose asymptotic form coincides with a solution to
the Einstein-Hilbert Lagrangian with a negative cosmological
constant. In this case, near the boundary of AdS we may use the
Einstein-Hilbert action in order to relate quantities in the bulk to
quantities in the field theory.

\subsection{The dual of the NCE}

Let us recall the relation between the extrinsic
curvature and the energy-momentum tensor in the case of the
Einstein-Hilbert action:
\begin{equation}
T^{\mu}_{\phantom{\mu}\nu} =
-\frac{1}{8\pi\,G_{d+1}}\(K^{\mu}_{\phantom{\mu}\nu}-K\,h^{\mu}_{\phantom{\mu}\nu}\).
\label{energy-momentum}
\end{equation}
As usual $h_{\mu\nu}$ is the induced metric on $r=const.$ and
$K_{\mu\nu}$ is its extrinsic curvature. In order to find the field theory dual of such
quantities a holographic
renormalization procedure is implemented. The renormalization
procedure of the energy-momentum tensor for the Einstein-Hilbert
action is  well-known (see for example~\cite{papa}).

Holographic renormalization requires to formally perform a transformation to Euclidean
signature in order to define the relation between the dual field
theory and the on-shell gravity action. Here we use all the
relations after transforming back to Lorenzian signature, assuming that such transformations can be performed without obstructions. The induced
energy-momentum tensor in the field theory is given by
\begin{equation}
\langle\ \(T_{\mu\nu}\)_{FT}\ \rangle    =\lim_{r \rightarrow
\infty}\(r^{d-2}\,T_{\mu\nu}\)_{ren}. \label{the EM tensor}
\end{equation}

Trace-reversing Eq.~(\ref{energy-momentum}) we obtain
\begin{equation}
K^{\mu}_{\phantom{\mu}\nu}=-8\pi\,G_{d+1}\(T^{\mu}_{\phantom{\mu}\nu}-\frac{T}{d-1}h^{\mu}_{\phantom{\mu}\nu}\).
\end{equation}

Then, we can transform this relation to the field theory and get a
relation in terms of the corresponding one-point functions:
\begin{eqnarray}
\langle    O_{\Theta^{t}_{\phantom{t}t}}\rangle
= \langle\ \(T^{t}_{\phantom{t}t}\)_{FT}\ \rangle -\frac{\langle T_{FT}\rangle}{d-1}
\cr
\langle O_{\Theta^{x}_{\phantom{x}x}}\rangle = -\langle\ \(T^{x}_{\phantom{x}x}\)_{FT}\ \rangle +\frac{\langle
T_{FT}\rangle }{d-1}
\label{thetas}
\end{eqnarray}

We assume that the energy-momentum of the field theory is of the
perfect fluid form $
\(T^{\mu}_{\phantom{\mu}\nu}\)_{FT}=diag(-\varepsilon,P,P,...) $ to
describe the thermodynamics of the field theory\footnote{ In the
case that the theory is conformal $\langle T_{CFT}\rangle=0$ and
Eqs.~(\ref{thetas}) simplify to $\langle
O_{\Theta^{t}_{\phantom{t}t}}\rangle     = -\varepsilon$ and
$\langle O_{\Theta^{x}_{\phantom{x}x}}\rangle =-P.$}. Then \bea
\lim_{r \rightarrow \infty}\( K^t_{\phantom{t}t}\)_{ren}&=&8\pi
G_{d+1}\(\varepsilon+\frac{\langle
T_{FT}\rangle}{d-1}\) \nonumber\\
\lim_{r \rightarrow \infty}\(K^x_{\phantom{x}x}\)_{ren}&=&8\pi
G_{d+1}\(-P+\frac{\langle T_{FT}\rangle}{d-1}\) \eea Recalling the
linear combination in Eq.~(\ref{eps}) we find that \be \langle
O_{{\cal S}}\rangle =\,\varepsilon+P=\lim_{r \rightarrow
\infty}\frac{\(
K^t_{\phantom{t}t}\)_{ren}-\(K^x_{\phantom{x}x}\)_{ren}}{8\pi
G_{d+1}}. \ee The entropy density of the field theory (which we
denote by $s$) is identified with the dual operator of ${\cal S}$ so
that
\begin{equation}
sT=\varepsilon+P=\langle    O_{{\cal S}}\rangle.
\end{equation}
The corresponding field in the bulk is $\sigma_t+\sigma_x$ and in
section~2 we have seen that it gives the entropy density at the
horizon. Here we have established the duality of the two
descriptions of entropy: The NCE in bulk and the field theory
entropy on the boundary.

\subsection{Asymptotic conditions}
\label{asymptotic}
We have required that asymptotically,  when
$r\rightarrow\infty$, Eq. ($\ref{energy-momentum}$) holds without
corrections to the energy-momentum one-point function. The condition
that this requirement holds depends on the type of the higher order
corrections to the Einstein-Hilbert Lagrangian and on the number of
space-time dimensions.  We wish to establish a criterion for the
validity of the asymptotic condition. If the condition is not
satisfied, we classify the higher derivative correction to the
action as a correction which modifies the theory on the boundary.
These interesting cases are left for a future study since their
analysis is more complicated.

In general, the energy-momentum tensor on a hypersurface is derived
from the action
\begin{equation}
T_{\mu\nu}=\frac{2}{\sqrt{-h}}\frac{\partial
K^{\alpha\beta}}{\partial \dot{h}^{\mu\nu}}\frac{\delta S}{\delta
K^{\alpha\beta}}.
\end{equation}
Since $K_{\beta\gamma}=\dot{h}_{\beta\gamma}/(2\,N)$ we can express the
field theory
energy-momentum tensor as follows
\begin{equation}
\langle\ \(T_{\mu\nu}\)_{FT}\ \rangle    =\lim_{r \rightarrow
\infty}\(\frac{r^{d-2}}{\sqrt{-g}}\,\frac{\delta S}{\delta
K^{\mu\nu}}\)_{ren}.  \label{1-point}
\end{equation}

In order to obtain one-point function of the
energy-momentum tensor we have to expand $\frac{\delta S}{\delta
K^{\mu\nu}}$ in the radial coordinate $r$ in the neighborhood of the
boundary $r \rightarrow \infty$ and according to Eq.~(\ref{1-point})
the relevant terms are only those that decay asymptotically as $1/r^{d-2}$. The expansion of the
metric for asymptotically AdS spacetimes (for the components which
are orthogonal to the radial direction) is given by
\begin{equation}
g_{\mu\nu}=g^{(0)}_{\mu\nu}r^2+g^{(2)}_{\mu\nu}+\frac{g^{(4)}_{\mu\nu}}{r^2}+\cdots+ \frac{g^{(d)}_{\mu\nu}}{r^{d-2}}+\cdots
\end{equation}

Let us expand a general action, whose leading term is the
Einstein-Hilbert action, in a parameter $\ell$ which has length
dimensions as following:
\begin{equation}
{\cal L}=R+\sum_{n=2}\,\ell^{2n} {\cal L}_{2n}.
\end{equation}
The  correction term ${\cal L}_{2n}$ has $2n$  derivatives
of the metric. Each ${\cal L}_{2n}$ has to be
decomposed on the hypersurface $r=const.$ according to the
Gauss-Codazzi-Ricci decomposition in Eq.~(\ref{Riem1}) and
can be expressed as a sum of products of components of the extrinsic and
intrinsic curvatures and their contractions . In this
decomposition it is important to include the appropriate
(generalized) Gibbons-Hawking surface terms so that the action is
written as a sum of bulk and surface contributions
$S=S_{bulk}+S_{``GH"}.$
The surface contributions are ``absorbed'' after the above rewriting of the action on a
hypersurface. The details of the action and its surface terms are
not important for the kind of dimensional analysis that we will
present here.

Now we would like to determine the leading radial dependence of the contribution
of ${\cal L}_{2n}$ to $\langle T_{\mu\nu} \rangle_{FT}$ and decide when
this contribution is relevant, namely, whether it contributes to terms of order $1/r^{d-2}$.

The asymptotic behavior of the extrinsic curvature is
\[K_{\mu\nu} \sim r^2.\] Any Riemann tensor that appears in the decomposition has no
derivatives with respect to $r$ since it describes the intrinsic
curvature of the hypersurface. Therefore
\[R_{\alpha\beta\gamma\delta} \sim r^4,\]
and due to our interest in the dependence of the leading term on $r$
we can count each appearance of the Riemann tensor as two factors of
the extrinsic curvature. Since $\langle T_{\mu\nu} \rangle_{FT}$ is
obtained as a variation of the action with respect to $K^{\mu\nu}$,
we are left with $2n-1$ factors of the extrinsic curvature
$K_{\mu\nu}$. (factors of the Riemann tensor are counted as
explained above).

Since each contribution of ${\cal L}_{2n}$ to $\langle T_{\mu\nu}
\rangle$ has to be a rank two covariant tensor, the $2n-1$ factors
of the extrinsic curvature have to be contracted with $2n-2$ inverse
metrics whose leading terms scale as
\begin{equation}
g^{\alpha\beta}\sim
\frac{g^{(0)}_{\alpha\beta}}{r^2}.
\end{equation}
Consequently, the leading terms in the contribution of  ${\cal
L}_{2n}$ to $\delta S/\delta K^{\alpha\beta}$ all scale as
$r^2$. We are interested in higher derivative actions for which Eq.
(\ref{energy-momentum}) is satisfied. This requires that all
contributions to order $1/r^{d-2}$ to this equation for a specific
higher derivative action cancel. Therefore this condition determines
some algebraic equations that the coefficients of the higher
derivative terms must satisfy. For example, if the higher derivative
action is $f(R)$ we require that $\lim\limits_{r \rightarrow
\infty}f'(R)=1$ (see section \ref{second example}). The algebraic
equations determine which higher derivative terms are ``relevant''
and which are ``irrelevant'' in the language of RG flow equations.

There are no higher derivative corrections to the one-point function
if all the corrections satisfy the algebraic equations. In this case
we can use the relation in Eq.~(\ref{energy-momentum}) for the
one-point function of the energy-momentum tensor. If the algebraic
conditions are not satisfied then Eq.~(\ref{energy-momentum}) should
be modified in order to obtain the correct one-point function.

\section{Renormalization group flow of effective couplings}

We wish to discuss the variation of the entropy density operator from the
horizon to the boundary and interpret this variation as a RG flow of
the effective gravitational couplings.

We have shown that
\begin{equation}
s T = \frac{1}{\kappa_{eff}^2}
\sqrt{-h}\(K^{t}_{\phantom{t}t}-K^{x}_{\phantom{x}x}\)|_{r\to\infty}
\label{conserved_quantity}
\end{equation}
when renormalized properly on the boundary at $r\to\infty$. The
quantity $\sqrt{-h}\(K^{t}_{\phantom{t}t}-K^{x}_{\phantom{x}x}\)$ is
also proportional to the entropy density operator on the ``second''
boundary, namely, the horizon. This was already realized
in~\cite{Buchel1,Liu1} for the Einstein-Hilbert action case (when
$s$ is the Bekenstein-Hawking entropy density).

For a black brane ansatz of the form  (\ref{brane}) we have
 \begin{eqnarray}
\sqrt{-h}\,K^{t}_{\phantom{t}t}&=&-\frac{g_{tt,r}}{2\,\sqrt{-g_{tt}g_{rr}}}\,(g_{xx})^{\frac{d-1}{2}},\\
\sqrt{-h}\,K^{x}_{\phantom{x}x}&=&\frac{g_{xx,r}}{2\,\sqrt{g_{xx}g_{rr}}}\,(g_{xx})^{\frac{d-2}{2}}\,\sqrt{-g_{tt}}.
\end{eqnarray}
The temperature at the horizon is given by
\begin{equation}
T=-\left.
\frac{g_{tt,r}}{4\,\pi\,\sqrt{-g_{tt}g_{rr}}}\right|_{r=r_h}\;\;\;,
\end{equation}
and the entropy density is
\begin{equation}
\left. s=\frac{2\,\pi}{\(\kappa_{eff}\)^2}\,
(g_{xx})^{\frac{d-1}{2}}\right|_{r=r_h}.
\end{equation}

The $tt$ component of the metric $g_{tt}$ vanishes at the horizon
and thus $K^{x}_{\phantom{x}x}$ vanishes there. So at the horizon
all the contribution to (\ref{conserved_quantity}) comes from
$K_{\phantom{t}t}^{t}$:
\begin{equation}
\left.
\frac{1}{\(\kappa_{eff}\)^2}\sqrt{-h}K_{\phantom{t}t}^{t}\right|_{r=r_h}=\,s\,T.
\end{equation}

In conclusion, it turns out that the difference $\sqrt{-h}\(
K^{t}_{\phantom{t}t}- K^{x}_{\phantom{x}x}\)$ gives $s T$ up to
$\(\kappa_{eff}\)^2$ on both boundaries - the AdS boundary and the
horizon. Our interpretation is, as we will see below in more detail,
that the effective coupling that flows from the horizon (UV) to the
boundary (IR) is related to the dependence of the entropy density
operator on the radial direction. According to the \ads
correspondence the entropy density of the black brane is equal to
the entropy density of the gauge theory on the boundary. This is a
direct consequence of the identification of the temperature of the
thermal state in the gauge theory with the Hawking temperature of
the black brane, and as a result the corresponding free energies on
both sides (see, for example,~\cite{aharony}).

Thus $s T$ is equal on both boundaries, so that the change in
$\kappa_{eff}$ between the two boundaries exactly cancels the change
in the entropy density operator $A$,
\begin{equation}
\label{field} A \equiv
\sqrt{-h}\(K^{t}_{\phantom{t}t}-K^{x}_{\phantom{x}x}\).
\end{equation}
Using the RG flow terminology, we consider rescaling with respect to
$r$. The effective coupling $\kappa_{eff}$ is a running coupling
between two scales. Its running is equal to the accumulated
rescaling of the corresponding operator $A$. Let us denote the
variation of the quantities between the two boundaries (scales) by
$\Delta$. Then we can write the RG flow equation
\begin{equation}
\label{RG flow}
2 \frac{\Delta \kappa_{eff}}{\kappa_{eff}}=\frac{\Delta A}{A}.
\end{equation}
In the rest of this section we give two examples to the RG flow of $\kappa_{eff}$
-- no RG flow for the Einstein-Hilbert action and a simple RG flow
for $f(R)$ Lagrangians. An additional non-trivial example is the RG
flow in the context of the leading order 8-derivative correction proportional to the fourth power of the Weyl tensor
to the type IIB black hole in $AdS_5$~\cite{dutta} which will be presented elsewhere.

\subsection{No RG flow for the Einstein-Hilbert action}

In the case of Einstein-Hilbert action the effective coupling is a
radial constant. From the gravity point of view it corresponds to
having a purely Einstein-Hilbert action in the UV. In this case we
do not expect any renormalization of Newton's constant in the IR.
From the AdS bulk perspective, no RG flow means that the entropy
density operator should not change with $r$, namely, that it is a
constant of the equations of motion. Here we show it explicitly
following~\cite{Buchel1}.

The Poincare symmetry of the brane geometry implies that the sources
of the brane should satisfy
\begin{equation}
T_{tt}+T_{xx}=0.
\end{equation}
From Einstein's equations it follows that  \begin{equation}    R_{tt}+R_{xx}=0,
\end{equation}
which is equivalent to
\begin{equation}
R_{\phantom{t}t}^{t}=R_{\phantom{x}x}^{x}.
\label{PoincareEinstein}
\end{equation}
(See \cite{springer} for more details).

For a brane ansatz of the form~(\ref{brane}) we can rewrite equation (\ref{PoincareEinstein}) as
\begin{equation}
\frac{1}{\sqrt{-g}}\frac{d}{dr}\(
\sqrt{-h}K^{t}_{\phantom{t}t}-\sqrt{-h}K^{x}_{\phantom{x}x}\)=0,
\label{conservation}
\end{equation}
so that $s\,T$ is conserved as we move from the horizon (UV) to the
boundary (IR).

\subsection{The flow in $f(R)$ Lagrangians}

\label{second example}
As a second example we would like to discuss
the NCE dual for Lagrangians that depend only on the Ricci scalar,
i.e. of the form ${\cal L}=f(R)$.

In order to find the relation of the extrinsic curvature to the
energy-momentum tensor let us introduce an auxiliary scalar field
$\phi$ and consider the following action:
\begin{equation}
\label{scalar action}
\widetilde{S}=\frac{1}{16\,\pi\,G_{d+1}}\int\!\! dr\,\int\!\! dt\,\int\!\!
d^{d-1}x \sqrt{-g}\,\left[f(\phi)+f'(\phi)\(R-\phi\)\right],
\end{equation}
which is equivalent on-shell to the original action for $f(R)$
(provided $f''(R)\neq 0$. For more details see, for example, \cite{scalar}). Then we can substitute
the decomposition of the Ricci scalar with respect to a hypersurface
$r=const.$
\begin{equation}
^{(d+1)}R=^{(d)}R+K^{\alpha\beta}K_{\alpha\beta}-K^{2}-2\,\nabla_{a}\(n^{b}\nabla_{b}n^{a}-n^{a}\nabla_{b}n^{b}\).
\end{equation}
The last term is canceled with the appropriate boundary term of the
Gibbons-Hawking type. The boundary term in $f(R)$ theories turns out
to be~\cite{scalar,scalar2} \[-2\int\!\! dt \int\!\!
d^{d}x\sqrt{-h}\,f'(R)\,K.\]

The energy momentum tensor for the action (\ref{scalar action}) is
given by
\begin{equation}
T^{\alpha\beta}=\frac{2}{\sqrt{-h}}\frac{\delta
\widetilde{S}}{\delta
\dot{h}_{\alpha\beta}}=\frac{2}{\sqrt{-h}}\frac{\partial
K_{\alpha\beta}}{\partial \dot{h}_{\alpha\beta}}\frac{\delta
\widetilde{S}}{\delta K_{\alpha\beta}},
\end{equation}

so that we get
\begin{equation} \label{EM tensor for f(R)}
T^{\alpha}_{\phantom{\alpha}\beta} =
-\frac{f'(R)}{8\,\pi\,G_{d+1}}\(K_{\phantom{\alpha}\beta}^{\alpha}-K\,h^{\alpha}_{\phantom{\alpha}\beta}\).
\end{equation}

Trace-reversing Eq.~(\ref{EM tensor for f(R)}) we obtain
\begin{equation} \label{tr-rv-f(R)}
K^{\alpha}_{\phantom{\alpha}\beta}=-\frac{8\pi\,G_{d+1}}{f'(R)}\(T^{\alpha}_{\phantom{\alpha}\beta}-\frac{T}{d-1}h^{\alpha}_{\phantom{\alpha}\beta}\).
\end{equation}

The entropy of the black brane is given by Wald's formula
(substituting, for example, in Eq.~(\ref{wald for static}))
\begin{equation}
S_{W}=\frac{A_{h}}{4\,G_{d+1}}\,\left. f'(R)\, \right|_{r=r_{h}}
\quad .
\end{equation}

The effective coupling in this case is
\begin{equation} \label{keff}
\(\kappa_{eff}\)^2=\frac{8\,\pi\,G_{d+1}}{f'(R)}.
\end{equation}
This effective coupling is in general $r$ dependent.
In this example the effective coupling changes as it
flows from the horizon to the boundary, which corresponds to a flow  from the UV to the IR. This
is consistent with the way that we have defined the NCE as a function of the radial coordinate.
In order for (\ref{energy-momentum}) to hold on the boundary we
require
\begin{equation}
\lim_{r \rightarrow \infty}f'(R)=1
\end{equation}
so that
\begin{equation}  \lim_{r \rightarrow
\infty}\(\kappa_{eff}\)^2=8\,\pi\,G_{d+1},
\end{equation}
which is its the value for the Einstein-Hilbert case.

The conservation law along the radial direction (\ref{conservation})
will no longer hold and,  in contrast to the Einstein-Hilbert case,
the equation will acquire a source term. The source term corresponds
to the fact that $\kappa_{eff}$ changes in the radial direction and
it is not a constant any more. For each value of $r$ it has a
different value.

This example illustrates how the RG flow equation (\ref{RG flow}) is
satisfied. From Eq. (\ref{tr-rv-f(R)}) and the definition of $A$
(\ref{field}) we observe that \be A=\frac{1}{f'(R)}A_{EH},\ee where
$A_{EH}$ is the value of $A$ in the case of Einstein-Hilbert
action\footnote{Any black hole solution of the Einstein equations
with a cosmological constant is also a solution of the $f(R)$
equations of motion~\cite{scalar}.} (no RG flow). Therefore
\begin{equation}
\frac{\Delta A}{A}=\left. \frac{1}{f'(R)}\right|_{r=r_h}-1.
\end{equation}
The same value is obtained also from $2 \Delta
\,\kappa_{eff}/\kappa_{eff}$ when $\kappa_{eff}$ was given in Eq.
(\ref{keff}). To conclude, in $f(R)$ Lagrangians we already observe
a non-trivial RG flow from the UV to the IR.

\subsection{Some comments on the general case}

For theories for which the algebraic equations discussed in section
\ref{asymptotic} are not satisfied it is more difficult to find the
explicit expression for the extrinsic curvature $K_{\mu\nu}$ in
terms of the energy-momentum tensor $T_{\mu\nu}$ on the boundary.
For instance, this relation was obtained only recently for the case
of a Gauss-Bonnet correction in~\cite{Liu2} and only perturbatively
in $\alpha'$.

We expect that the a mapping of the entropy from
the bulk to the boundary will exist also for any general Lagrangian of the form (\ref{theaction}). The correction terms could be analyzed according to their relevance or irrelevance as in the standard RG analysis. We expect relations between the  conformal invariance (or covariance) properties of operators in the field theory and the nature of the RG flow of their duals in the bulk.

\section{Summary and conclusions}

In this paper we have expressed the Noether charge entropy density
of a black brane in AdS space in terms of local operators
in the AdS space bulk. This allowed us to determine the
field theory dual of the Noether charge entropy for theories that
asymptote to Einstein theory. We have defined such theories as
theories in which the relation of the extrinsic curvature to the
energy-momentum tensor on the boundary is the same as for the
Einstein-Hilbert action (\ref{energy-momentum}) for relevant terms,
terms that decay asymptotically no faster than $1/r^{d-2}$. Such
asymptotic dependence is dictated by the holographic renormalization
procedure that determines the one-point functions of the dual
operators. The latter are the energy density $\varepsilon$ and
pressure $P$ of the field theory, so that the entropy density of the
field theory is obtained as the combination
$$ sT=\varepsilon+P. $$

We have interpreted the variation of the entropy density operator
from the horizon to the boundary as due to the renormalization of
the effective gravitational couplings when they flow from the UV to
the IR. We have found that the RG flow equation for the effective
coupling is related to the scaling of the entropy density operator:
$$
2 \, \frac{\Delta \kappa_{eff}}{\kappa_{eff}}=\frac{\Delta A}{A},
$$
where
\[A\equiv \sqrt{-h}\(K^{t}_{\phantom{t}t}-K^{x}_{\phantom{x}x}\).\]

For the Einstein-Hilbert theory we found that there is no RG flow.
In the case of $f(R)$ theories we have found the RG flow explicitly
as
$$
2 \, \frac{\Delta
\kappa_{eff}}{\kappa_{eff}}=\frac{1}{f'(R)}\biggr|_{r=r_h}-1.
$$
When corrections introduce new relevant terms and Eq.
(\ref{energy-momentum}) is not satisfied, the situation is more
complicated and interpreted as a modification of the theory on the
boundary. We leave such cases for future study.

Our results support the association of the NCE density operator with an effective
gravitational coupling constant, an association that was introduced for the
first time in~\cite{first}.

\section{Acknowledgments}

We thank Ofer Aharony, Alex Buchel, Suvankar Dutta, Valeri Frolov,
Ingmar Kanitscheider, Kostas Skenderis, Marika Taylor and Amos Yarom
for discussions. We thank Joey Medved for comments on the
manuscript.

The research of RB was supported by The Israel Science Foundation
grant no 470/06. DG thanks the National Sciences and Engineering
Research Council of Canada for the financial support.

\end{document}